\documentclass{article}
\usepackage[dvips]{graphicx}
\newtheorem{fig}{Figure}

\newtheorem{tab}{Table}

\def\ligne#1{\hbox to \hsize{#1}}

\def\leurre{\noindent\leftskip0pt\small\baselineskip 10pt}
\def\grostrait{\ligne{\vrule height 1pt depth 1pt width \hsize}}
\def\demitrait{\ligne{\vrule height 0.5pt depth 0.5pt width \hsize}}

\title{
Bacteria inspired patterns grown with
hyperbolic cellular automata}
\author{Maurice Margenstern\\
Laboratoire d'Informatique Th\'eorique et Appliqu\'ee, EA 3097,\\
        Universit\'e Paul Verlaine $-$ Metz, UFR-MIM, \\
        and CNRS, LORIA,\\
        \^Ile du Saulcy, 57045 Metz Cedex, France\\
{\it e-mail}: {\tt margens@univ-metz.fr}
}

\begin{document}
\maketitle

\begin{abstract}
In this paper we give three examples of expending patterns defined by
hyperbolic cellular automata whose growth seems to be very similar to the 
growth of colonies of bacteria.
\end{abstract}

\section{introduction}
\label{intro}
   A very intriguing phenomenon of diffusion is given by the growth of colonies
of bacteria, see~\cite{benjacob}. As 
explained by Professor Ben-Jacob, very surprising
structures can be obtained by putting such colonies in very severe conditions, 
see Figures~\ref{bacteria} and~\ref{compare_bacteria}. 
This gives a striking power of adaptability of these colonies. These 
experimental data are comforted by the discovery of bacteria in almost 
every possible hard 
conditions as geysers, ocean fathoms, core of the earth and even atomic piles.
In the introduction of~\cite{benjacob}, Professor Ben-Jacob says:
\vskip 7pt
{\leftskip 20pt\rightskip 20pt\small
Eons before humans, bacteria inhabited a very different Earth.
As the earliest life form they devised ways to counter the spontaneous
course of increasing entropy and convert high-entropy, inorganic substances
into low entropy, organic molecules...

To change environmental hazards, bacteria resort to a wide range of
cooperative strategies...

They collectively glean information from the environment, communicate,
distribute tasks, perform distributed information processing and learn
from past experience.
\par}
\vskip 7pt

   In many cases, their growth on plates used by microbiologists to study them
constitute figures with a more or less fractal symmetry. This may address to
hyperbolic geometry. This is why we tried the other way: let us start from
hyperbolic geometry, in fact from an appropriate tiling of the hyperbolic plane
and try to simulate the observed growth.

   In Section~\ref{geom} we give the needed information for the reader about 
what to know about hyperbolic geometry in order to understand how our grid is 
obtained and to see how cellular automata are implemented in this context.
In Section~\ref{simul}, we see how to proceed to the simulations indicated
in the abstract. We conclude in Section~\ref{conclusion} with ideas about
possible continuations.

\vskip 10pt
\vtop{
\ligne{\hfill
\scalebox{0.275}{\includegraphics{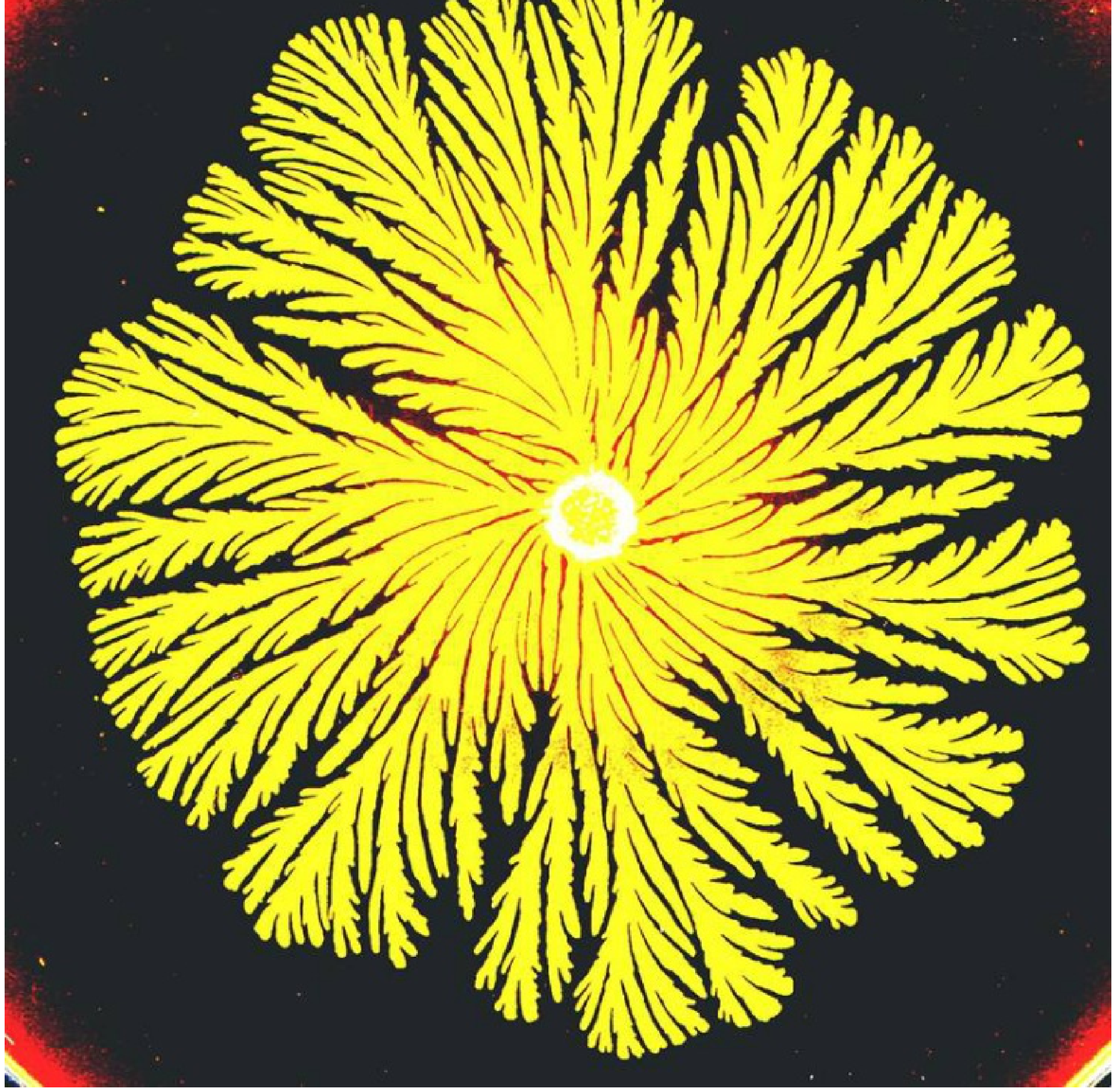}}
\scalebox{0.275}{\includegraphics{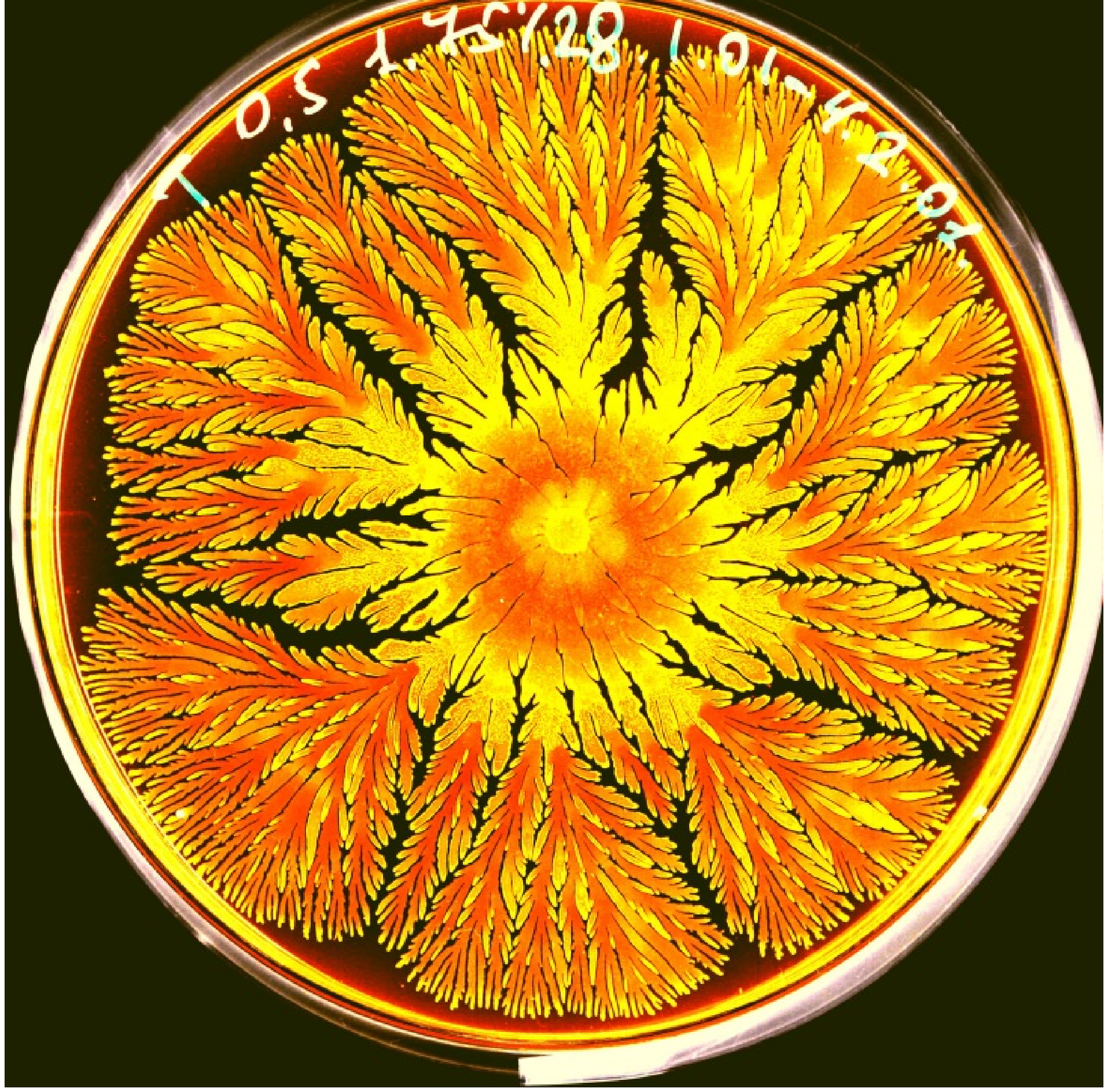}}
\hfill}
\vspace{-70pt}
\begin{fig}\label{bacteria}
\leurre
Examples of growth of bacteria colonies. Pictures by courtesy of Professor 
Ben-Jacob.
\end{fig}
}
\vspace{-30pt}
\section{A triangular tiling of the hyperbolic plane} 
\label{geom}
   In this section, we provide the reader with the minimal material which allows 
the paper to be self-contained. Sub-section~\ref{introgeom} is a very short
introduction to hyperbolic geometry and to the Poincar\'e's disc model which is
intensively used in the illustrations of the paper. Then, 
Sub-section~\ref{grid} indicates which grid we consider for the simulations we
propose. Then, in Sub-section~\ref{HCA} we introduce cellular automata adapted
to this grid and the discussion we have there will lead to the various examples
of simulation dealt with in Section~\ref{simul}.

\subsection{Hyperbolic geometry}
\label{introgeom}

Hyperbolic geometry appeared in the first half of the
19$^{\rm th}$ century, proving the independence of the parallel
axiom of Euclidean geometry. Models were devised in the second
half of the 19$^{\rm th}$ century and we shall use here one of the
most popular ones, Poincar\'e's disc. This model is represented by
Figure~\ref{poincare_disc}.

Inside the open disc represented in the figure we have the points of the 
hyperbolic plane. Note that by definition, the points on the border of the disc 
do not belong to the hyperbolic plane. However, these points an important role
in this geometry and are called {\bf points at infinity}. Lines are trace of 
diameters or circles orthogonal to the border of the disc, e.g. the line~$m$. 
In this model, two lines which meet in the open disc are called {\bf secant}
and two lines which meet at infinity, {\it i.e.} at a point at infinity
are called {\bf parallel}. In the figure, we can see a line~$s$ through 
the point~$A$ which cuts~$m$. Now, we can see that two lines pass through~$A$
which are parallel to~$m$: $p$ and~$q$. They touch~$m$ in the model at~$P$ and~$Q$
respectively which are points at infinity. At last, and not the least: 
the line~$n$ also passes through~$A$ without cutting~$m$,
neither inside the disc nor outside it. This line is called {\bf non-secant}.

\vskip 10pt
\vtop{
\ligne{\hfill
\scalebox{1.00}{\includegraphics{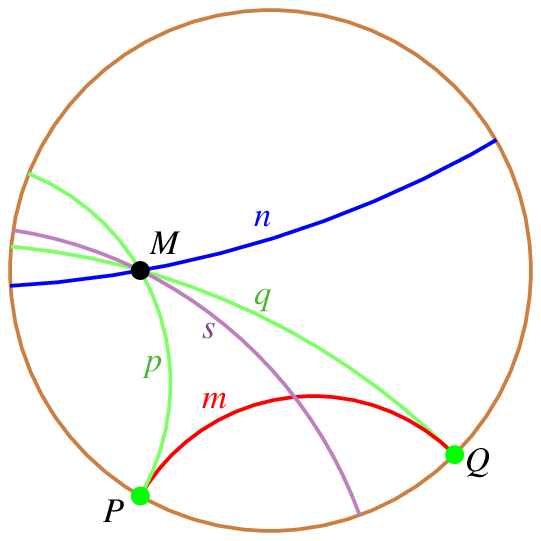}}
\hfill}
\begin{fig}\label{poincare_disc}
\leurre
Poincar\'e's disc model.
\end{fig}
}

\subsection{The grid of our simulations}
\label{grid}

From a famous theorem established by Poincar\'e in the late 19$^{\rm th}$ century,
it is known that there are infinitely many tilings in the hyperbolic plane, 
each one generated by the reflection of a convex polygon~$P$ in its sides and, 
recursively, in the reflection of the images in their sides, provided that the 
number~$p$ of sides of~$P$ and the number~$q$ of copies of~$P$ which can be put 
around a point~$A$ and exactly covering a neighbourhood of~$A$ without overlapping 
satisfy the relation:
\hbox{$\displaystyle{1\over p}+\displaystyle{1\over q}<\displaystyle{1\over2}$}.
The numbers $p$ and~$q$ characterize the tiling which is denoted $\{p,q\}$ and
the condition says that the considered polygons live in the hyperbolic plane. Note
that the three tilings of the Euclidean plane which can be defined up to similarities
can be characterized by the relation obtained by replacing~$<$ with~$=$ in the above
expression. We get, in this way, $\{4,4\}$ for the square, $\{3,6\}$ for the 
equilateral triangle and $\{6,3\}$ for the regular hexagon.

   In the paper, we shall focus our attention on one of the simplest tilings 
which can be defined in this way in the hyperbolic plane: $\{7,3\}$. We call this
tiling the {\bf heptagrid} which is illustrated by Figure~\ref{tiling_73}.
On the right-hand side of the figure, we can see the tree which is in bijection
with an angular sector, a basic structure of the heptagrid, also see
Figure~\ref{eclate_73}. This property is the basis of a very efficient navigation
tool to locate tiles in the heptagrid.

   We refer the reader to~\cite{mmbook1,mmbook2} for a detailed analyses and
detailed explanations of these tools.

In the book, it is also indicated how to derive from the heptagrid an infinite
family of tilings. We proceed to this point now.

\vskip 10pt
\vtop{
\ligne{\hfill
\scalebox{0.40}{\includegraphics{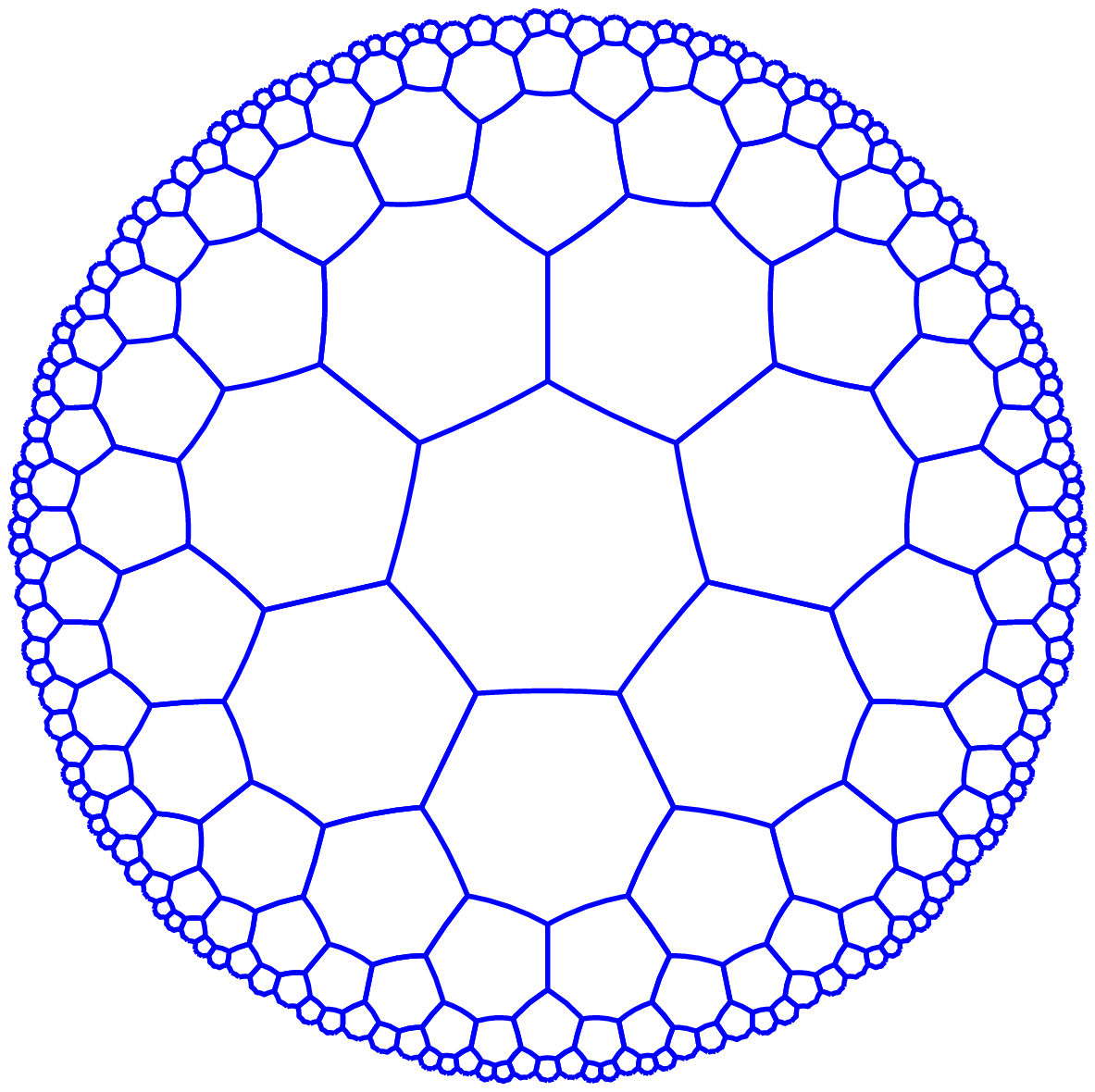}}
\scalebox{1.00}{\includegraphics{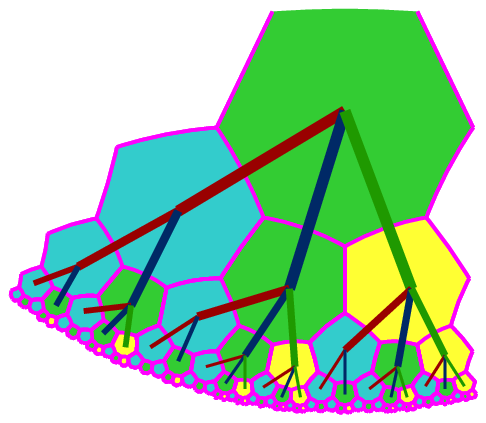}}
\hfill}
\begin{fig}\label{tiling_73}
\leurre
The heptagrid. On the right-hand side: the key structure to explore the tiling.
\end{fig}
}

\vskip 10pt
\vtop{
\ligne{\hfill
\scalebox{0.30}{\includegraphics{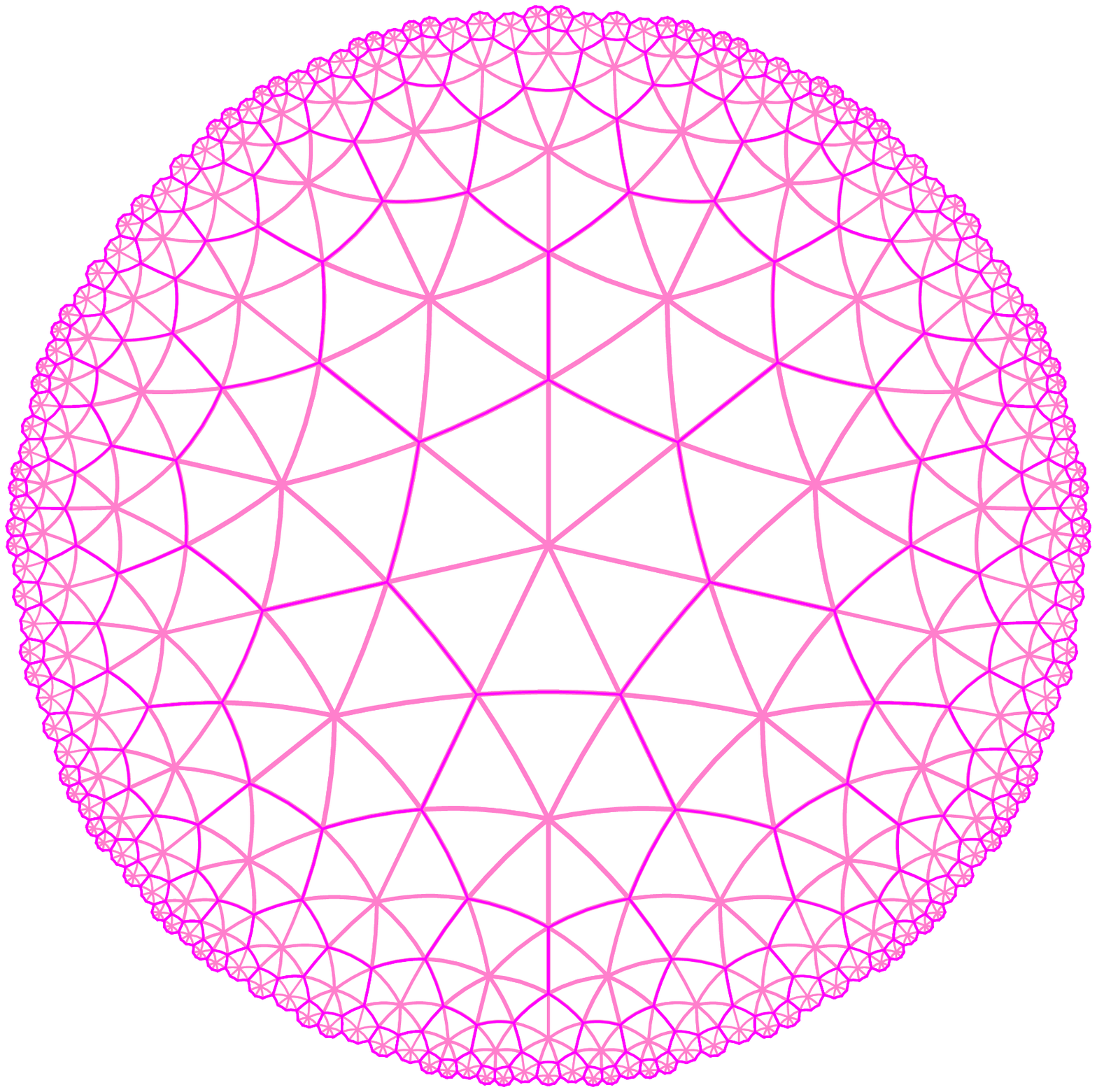}}
\scalebox{0.30}{\includegraphics{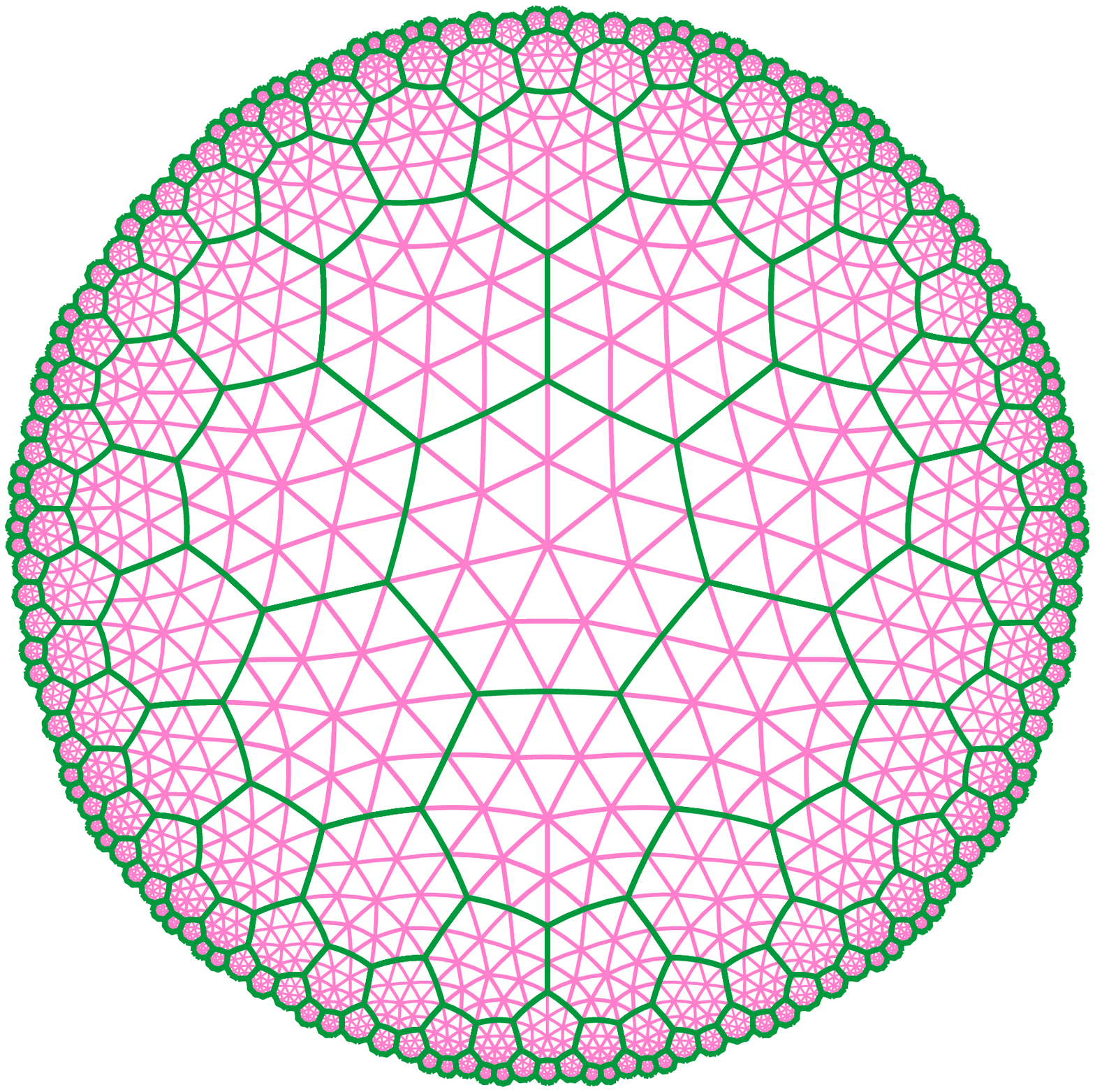}}
\hfill}
\begin{fig}\label{subtilings}
\leurre
The subtilings of the heptagrid. On the left-hand side: generation~$1$;
on the right-hand side, how generation~$2$, the {\bf heptatrigrid},
is built from generation~$1$.
\end{fig}
}

   The idea is to  split each tile of the heptagrid into seven triangles
whose vertices are the centre of the tile and the end-points of its edges.
Then, we split the triangles into four triangles by taking the mid-points
of the previous triangles: the new triangles are defined by a vertex of the
previous triangles and the mid-points of the edges of the previous triangle
which meet at the vertex. This defines generation~1 and generation~2
of the process, see Figure~\ref{subtilings}. We can go on the process 
inductively: the generation $n$+1 is obtained from the generation~$n$ as 
generation~2 is obtained from generation~1. In~\cite{mmbook2}, this process
is defined up to infinity, giving the possibility to define coordinates for
the points of the hyperbolic plane.

However, in this paper we shall focus on generation~2 only. Let us call it 
the {\bf second triangular heptagrid}, {\bf heptatrigrid} for short.

\vskip 10pt
\vtop{
\ligne{\hfill
\scalebox{0.40}{\includegraphics{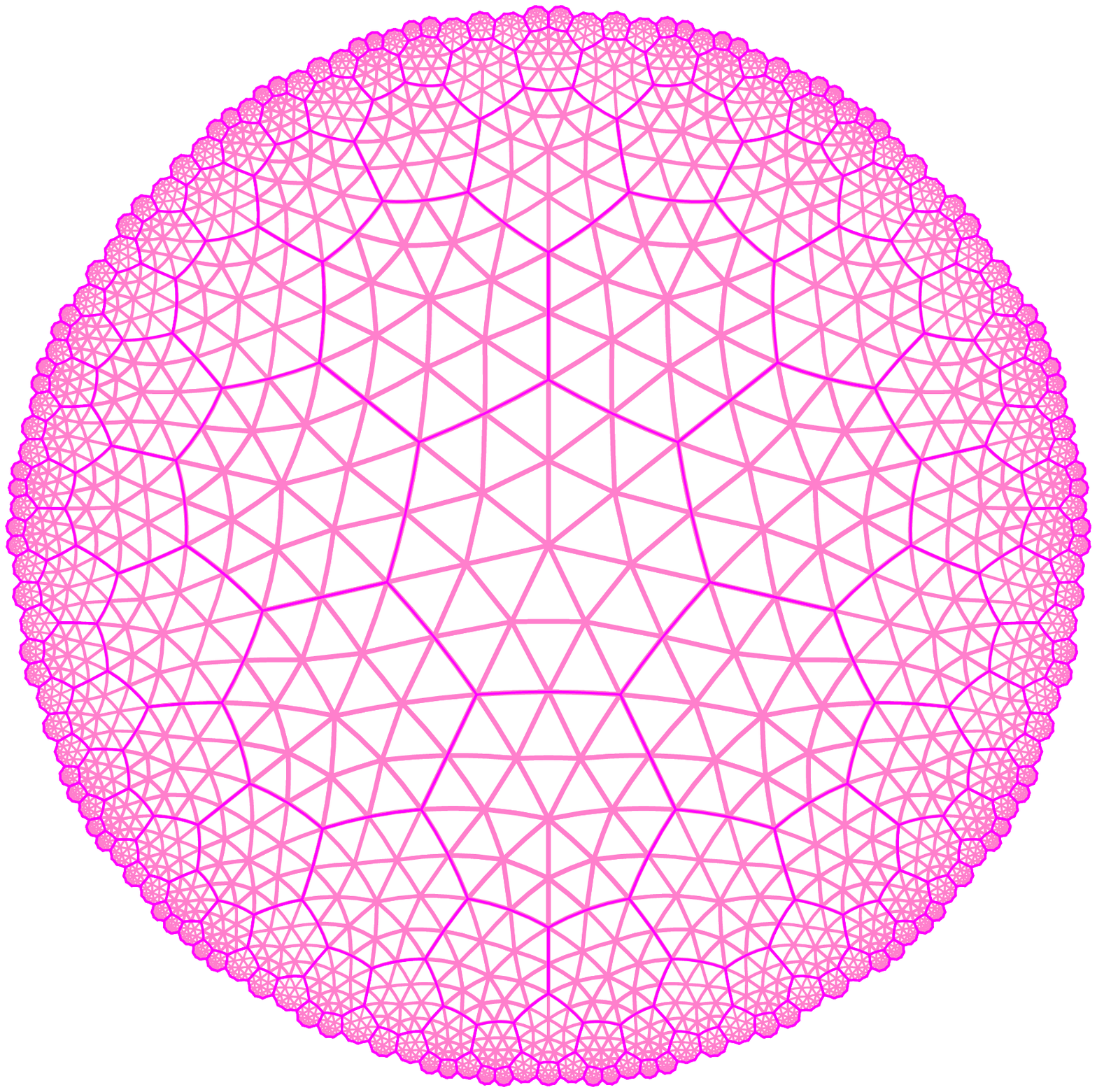}}
\hfill}
\begin{fig}\label{heptatrigrid}
\leurre
The heptatrigrid.
\end{fig}
}

\subsection{Cellular automata on the hyperbolic triangular grid}
\label{HCA}

   Now, we turn to the implementation of a cellular automaton in the heptatrigrid.

   Remember that such an automaton consists in a finite automaton~$A$ attached
to each 2-triangle of the heptatrigrid. A cell of the finite automaton
consists of~$A$ and a 2-triangle which is the {\bf support} of the cell.
The neighbours of a cell~$c$ with $T$~as its support have,as their supports, the
2-triangles which share a side with~$T$. 

Figure~\ref{eclate_73} indicates the 
basic elements of the location of a triangle.
A coordinate is defined by four numbers in the format $(\sigma,\nu,\tau,\pi)$: 
$\sigma$~is the number of the sector in which the triangle lies; $\nu$~is the
number of the heptagon of the sector in which the triangle lies; $\tau$~is
the number in \hbox{[1..7]} of the generation~1 triangle in which the triangle lies;
and in this triangle, $\pi$~is the number of the triangle itself. Later on,
$\tau$~will be called the {\bf slice} and $\pi$~will be called the {\bf place}.
Note that the central heptagon has the coordinate $(0,0)$ as it belongs to
no sector.

   The numbering of the generation~1 triangle, we say later {\bf 1-triangle},
is defined by the number of the side of the heptagon on which the
1-triangle is built, which explains why this numbers is in~\hbox{[1..7]}. 
For the central heptagon, side~1 is fixed once for all
and the other sides are numbered by counter-clockwise turning around the tile.
The side~$i$ touches the heading heptagon of the sector~$i$, with $i$~in~[1..7].
For the other heptagons, side~1 is shared by the father of the heptagon in the
tree. Note that we consider that the father of the root of the tree is the central
cell.

\vskip 10pt
\vtop{
\ligne{\hfill
\scalebox{1.00}{\includegraphics{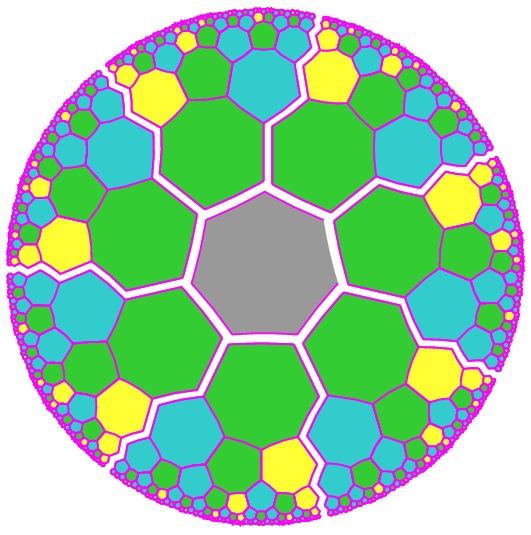}}
\scalebox{0.40}{\includegraphics{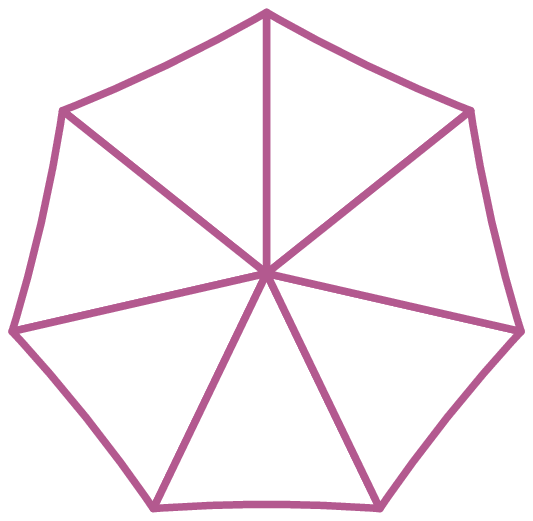}}
\hskip-75pt
\raise-15pt\hbox{\scalebox{0.35}{\includegraphics{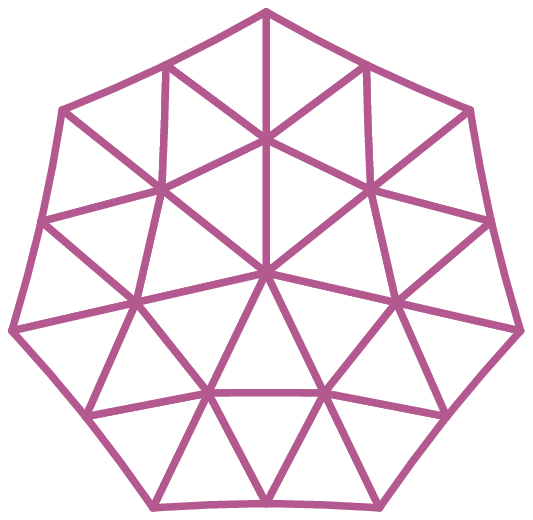}}}
\hfill}
\begin{fig}\label{eclate_73}
\leurre
From the heptagrid to the second triangular heptagrid, the heptatrigrid. 
\end{fig}
}

In each 1-triangle, we have four triangles of generation~2, we call them
{\bf 2-triangles}. The four 2-triangles of a 1-triangle are numbered from
0~to~3. First, we number the vertices of a 1-triangle~$T$ from 0~to~2: 2~is the
centre of the heptagon, 0~and~1 are the vertices of the side of the heptagon
defining~$T$. Following the counter-clockwise orientation, 0~comes before~1.
Now, the number of a vertex of~$T$ is the number of the 2-triangle which
possesses this vertex. Accordingly, 3~is the number of the 2-triangle whose
vertices are the mid-points of the edges of~$T$. This numbering can be repeated 
for any further generation, see~\cite{mmbook2}. This  numbering has interesting
properties. The interested reader is referred to~\cite{mmbook2} for more
information.

   Now, to implement cellular automata, we have to compute the coordinates of
the neighbours of a cell~$c$ from the coordinate of~$c$ itself. Let~$T$ be the
2-triangle which supports the cell. Number the neighbours of~$c$ from~0 to~3, 
$c$~itself being neighbour~0. For the other numbers, the neighbour~$i$ of~$c$ 
is the 2-triangle which shares the side~$i$ of~$T$. Let $(\sigma,\nu,\tau,\pi)$
be the coordinate of~$T$, the support of~$c$. The coordinates of the neighbours
of~$T$ are given by Table~\ref{coord1}. 

   As can be seen in the table, each 2-triangle has at least one neighbour which
is in the same heptagon and in the same slice of heptagon. Note that a 2-triangle
with place~3 has all its neighbours in the same slice of the same heptagon.
A 2-triangle with place~2 has all its neighbours in the same heptagon, but two
neighbours are in the slices which are adjacent to its own one. This is
indicated by the expressions $\tau\ominus1$ and $\tau\oplus1$. As
the number of a slice is in~$[1..7]$, subtracting~1 from~1 gives~7 
and adding~1 to~7 gives~1. For $\tau\in[2..7]$, \hbox{$\tau\ominus1=\tau$$-$1},
and for $\tau\in[1..6]$, \hbox{$\tau\oplus1=\tau$+1}. 
For the 2-triangles with place~0 and~1, the computation of the coordinates of
their neighbours is more complex. Indeed, in each case, one of the neighbours
do not belong to the same heptagon~$H$, but to a heptagon~$K$ neighbouring~$H$.
This changes the value of~$\nu$ and it may also change the values of~$\sigma$
and~$\tau$. This is indicated in Table~\ref{coord1} by the expressions 
$s(\sigma,\nu\,\tau)$, $v(\tau,\nu)$ and $t(\nu,\tau)$.

The computation of these expressions can be found in~\cite{mmbook1,mmbook2}
but we repeat them for the self-containedness of the paper.
The main point is the computation of~$v(\tau,\nu)$ as it may involve
auxiliary functions, namely $f(\nu)$, the coordinate of the father of the
tile~$\nu$ in its tree, $\sigma(\nu)$, the coordinate of the {\bf preferred son} 
of~$\nu$ and $st(\nu)$, the {\bf status} of~$\nu$. As mentioned 
from~\cite{mmkmTCS,mmJUCSii}, the tree which we consider has two kinds of nodes:
black and white ones with two and three sons respectively. The sons can be
deduced from the node by the following rules \hbox{$B\rightarrow B_*W$}
and \hbox{$W\rightarrow BW_*W$} in easy notations, where the star indicates the
place of the preferred son. We assume here that the functions $f(\nu)$ 
and~$\sigma(\nu)$ are known. Their computation is efficient but it requires
notions which we have not mentioned here. We refer the reader 
to~\cite{mmJUCSii,mmbook1,mmbook2} for additional information.

\def\lignetab #1 #2 #3 #4 #5 {%
\ligne{\hfill\hbox to 300pt{
\hbox to 80pt{\hfill#1\hfill}
\hbox to 50pt{\hfill#2\hfill}
\hbox to 50pt{\hfill#3\hfill}
\hbox to 50pt{\hfill#4\hfill}
\hbox to 50pt{\hfill#5\hfill}
}
\hfill}
}
\vtop{
\begin{tab}\label{coord1}
\leurre 
Coordinates of the neighbours of a $2$-triangle~$T$ in terms of the coordinates 
of~$T$.
\end{tab}
\vspace{-8pt}
\grostrait
\lignetab {neighbour} {sector} {number} {slice} {place}
\vspace{-3pt}
\demitrait
\lignetab {0} {$\sigma$} {$\nu$} {$\tau$} 0
\lignetab {1} {$\sigma$} {$\nu$} {$\tau$} 3
\lignetab {2} {$s(\sigma,\nu,\tau)$} {$v(\tau,\nu)$} {$t(\nu,\tau)$} 1
\lignetab {3} {$\sigma$} {$\nu$} {$\tau\ominus1$} 1
\vspace{-3pt}
\demitrait
\lignetab {0} {$\sigma$} {$\nu$} {$\tau$} 1
\lignetab {1} {$s(\sigma,\nu\tau)$} {$v(\tau,\nu)$} {$t(\nu,\tau)$} 0
\lignetab {2} {$\sigma$} {$\nu$} {$\tau$} 3
\lignetab {3} {$\sigma$} {$\nu$} {$\tau\oplus1$} 0
\vspace{-3pt}
\demitrait
\lignetab {0} {$\sigma$} {$\nu$} {$\tau$} 2
\lignetab {1} {$\sigma$} {$\nu$} {$\tau\ominus1$} 2
\lignetab {2} {$\sigma$} {$\nu$} {$\tau\oplus1$} 2
\lignetab {3} {$\sigma$} {$\nu$} {$\tau$} 3
\vspace{-3pt}
\demitrait
\lignetab {0} {$\sigma$} {$\nu$} {$\tau$} 3
\lignetab {1} {$\sigma$} {$\nu$} {$\tau$} 0
\lignetab {2} {$\sigma$} {$\nu$} {$\tau$} 1
\lignetab {3} {$\sigma$} {$\nu$} {$\tau$} 2
\vspace{-3pt}
\demitrait
\vskip 7pt
}
\vskip 5pt
Clearly,
the coordinate of a neighbour~$N$ of a tile~$T$ with coordinate~$\nu$ depends on
the side~$\tau$ shared by~$T$ and~$N$. Now, notice that this side numbered by~$\tau$
in~$T$ does not receive the same number in~$N$ and we shall say that $N$~is
the neighbour~$\tau$ of~$T$. The correspondence between these
numbers gives the value of the function $t(\nu,\tau)$ and, for completeness,
we give it in Table~\ref{coord2}. Note that the sides of the central cell
are all numbered by~1{} in its neighbours. For the other cells, the correspondence
depends on the status of~$T$ and it may also depend on that of~$N$.
Side~7 is always the side shared by a neighbour which is on the same level of the
tree, even when there is a change of tree by the change of sector. If~$T$ is
black, its side~7 is numbered~2 on the other side. If~$T$ is white, the number
of its side~7{} in the other neighbour~$N$ depends on the status of~$N$ as indicated
in the table.

   From this table, we can indicate the values of $v(\tau,\nu)$ which are given
in Table~\ref{coord3}. The basic point is that $v(1,\nu)$ for the heptagon~$H$
defined by~$\nu$ is always $f(\nu)$, as its neighbour~1 is the father of~$H$. 
Similarly, we have that $v(4,\nu)$ is always $\sigma(\nu)$ as neighbour~4 is 
the preferred son of~$H$, regardless of the status of~$H$. Note that in the case 
of a black heptagon~$H$ on the leftmost branch of the tree, two of its neighbours 
belong to the other tree on this side of the sector of~$H$: neighbours~2 and~3.
Neighbour~2 is still \hbox{$\nu$$-$1} and, consequently, neighbour~3 
is the rightmost son neighbour~2, hence it is 
\hbox{$\sigma(\nu$$-$$1)$+$1=\sigma(\nu)$$-$$1$}. 
A symmetrical remark holds for a white~$H$ standing on the rightmost branch of 
the tree: neighbours~6 and~7 belong to another tree, the one which spans the 
other sector than that of~$H$. Now, this time, neighbour~6 is 
numbered \hbox{$\nu$+1} and so, as neighbour~7 the father of neighbour~6,
neighbour~7 is numbered \hbox{$f(\nu$+$1)=f(\nu)$+1}. At last, the root, 
which is a 
white node, belongs to both the left- and the rightmost branches of the tree. 
This is why it has a specific profile, different from both a standard white 
node and from a node on the rightmost branch of the tree. 

\def\lignetabii #1 #2 #3 #4 {%
\ligne{\hfill\hbox to 300pt{
\hbox to 50pt{\hfill#1\hfill}
\hbox to 50pt{\hfill#2\hfill}
\hbox to 50pt{\hfill#3\hfill}
\hbox to 50pt{\hfill#4\hfill}
}
\hfill}
}
\vskip -10pt
\vtop{
\begin{tab}\label{coord2}
\leurre 
Correspondence between the numbers of a side shared by two heptagons,
$H$ and~$K$. Note that if $H$~is white, the other number of side~$1$
may be~$4$ or~$5$ when $K$~is white and that it is always~$5$ when
$K$~is black.
\end{tab}
\vspace{-8pt}
\grostrait
\ligne{\hfill\hbox to 100pt{\hfill black $H$\hfill}
\hfill
\hbox to 100pt{\hfill white $H$\hfill}\hfill}
\lignetabii {in $H$} {in $K$} {in $H$} {in $K$}
\vspace{-3pt}
\demitrait
\lignetabii 1 {3$^{wK}, $4$^{bK}$} 1 {4$^{wK}$, 5} 
\lignetabii 2      6     2     7 
\lignetabii 3      7     3     1
\lignetabii 4      1     4     1 
\lignetabii 5      1     5     1 
\lignetabii 6      2     6     2 
\lignetabii 7      2     7 {2$^{wK}$, 3$^{bK}$} 
\demitrait
\vskip 7pt
}

   It remains to indicate that in the case of a heptagon~$H$ which is on the
left- or the rightmost branch, it is easy to define the number of the sector
to which belongs the neighbours which do not belong to the tree of~$H$.
Indeed, let $\sigma$~be the number of the sector in which $H$~lies. If $H$~is 
a black node, its neighbours~2 and~3 are in the sector 
$\sigma\oplus1$. If $H$~is a white node, its neighbours~6 and~7
are in the sector $\sigma\ominus1$. Note that for the root of the sector~$\sigma$, 
its neighbour~2 is in the sector $\sigma\oplus1$ and its neighbour~1 is the central
cell which is outside all the sectors.

   To conclude with this section, let us remember that cellular automaton
have been implemented in several grids of the hyperbolic plane. The complexity
classes of these cellular automata have been investigated leading to very 
surprising
results. Several universal cellular automata also have been implemented in these
grids. We refer the reader to~\cite{mmbook2,mmLNCScmc11} for more information
and more references.

\def\lignetabiii #1 #2 #3 #4 #5 #6 #7 #8 {%
\ligne{\hfill\hbox to 340pt{
\hbox to 40pt{\hfill#1\hfill}
\hbox to 30pt{\hfill#2\hfill}
\hbox to 40pt{\hfill#3\hfill}
\hbox to 30pt{\hfill#4\hfill}
\hbox to 40pt{\hfill#5\hfill}
\hbox to 40pt{\hfill#6\hfill}
\hbox to 40pt{\hfill#7\hfill}
\hbox to 40pt{\hfill#8\hfill}
}
\hfill}
}

\vtop{
\begin{tab}\label{coord3}
\leurre 
The values of $v(\tau,\nu)$.
\end{tab}
\vspace{-8pt}
\grostrait
\lignetabiii {$\tau$} 1 2 3 4 5 6 7
\vspace{-3pt}
\demitrait
\lignetabiii {black} {$f(\nu)$} {$f(\nu)$$-$1} {$\nu$$-$1} {$\sigma(\nu)$} 
{$\sigma(\nu)$+1} {$\sigma(\nu)$+2} {$\nu$+1} 
\lignetabiii {left} {$f(\nu)$} {$\nu$$-$1} 
                    {$\sigma(\nu)$$-$1} {$\sigma(\nu)$} 
{$\sigma(\nu)$+1} {$\sigma(\nu)$+2} {$\nu$+1} 
\lignetabiii {white} {$f(\nu)$} {$\nu$$-$1} {$\sigma(\nu)$$-$1} {$\sigma(\nu)$} 
{$\sigma(\nu)$+1} {$\sigma(\nu)$+2} {$\nu$+1} 
\lignetabiii {right} {$f(\nu)$} {$\nu$$-$1} {$\sigma(\nu)$$-$1} {$\sigma(\nu)$} 
{$\sigma(\nu)$+1} {$\nu$+1} {$f(\nu)$+$1$} 
\lignetabiii {root} 0 1 {$\sigma(\nu)$$-$1} {$\sigma(\nu)$} {$\sigma(\nu)$+1} 
{$\nu$+1} 1
\vspace{-3pt}
\vspace{-3pt}
\demitrait
\vskip 3pt
\noindent
{\small\sc Note}: \leurre {\rm left} denotes the leftmost branch of the tree,
{\rm right} denotes its rightmost one. 
\vskip 7pt
}
\vskip 5pt
\section{The simulations}
\label{simul}

   Now, we have the tools to implement cellular automata in the 2-triangles.
In Sub-section~\ref{propa_tree}, we shall look at the implementation of
a cellular automata which propagates the tree structure of the heptagrid.
The result, illustrated by Figure~\ref{propa_tree_fig} convinced us that
we could try to simulate colonies of bacteria. We propose three of them
which are examined in Sub-section~\ref{othersimul} and which differ by
the number of states of the cellular automaton which is used for the
simulation.
 
\subsection{First simulation: the propagation of the tree structure}
\label{propa_tree}

   The tree structure of the heptagrid can be implemented by cellular automata
on this grid: this was illustrated in~\cite{mmTCS4sthepta} in order to give
a toy example of a cellular automaton on this grid.

   We can do the same here and Figure~\ref{propa_tree_fig} gives the 36$^{\rm th}$
step of execution of this automaton starting from an initial configuration
in which the seven 2-triangles of place~2 of a heptagon are in the same state,
in red in the figure: we call this the core-2 configuration. As we can see, 
the automaton has a non-small number of states: 18~of them. 
In~\cite{mmTCS4sthepta}, we had 5 states only. In fact, it is possible
to have 4~states in the case of the heptagrid if we do not need to differentiate
the two white sons of a white node. We need much more states here as we wish
to diffuse the structure of the tree with its two types of rules. For programming
reasons, it was easier to program the automaton by implementing the following
strategy: when the automaton enters a heptagon, it goes as soon as possible to
the 2-triangles with place~2. There, by a counting process, it determines the
directions of the sons from the direction of the father which is the direction
from which the automaton entered the cell.
   
  The way the automaton is working can be seen as an animation on the slides
which are deposited on~\cite{mmsite}.

\vskip 10pt
\vtop{
\ligne{\hfill
\scalebox{0.30}{\includegraphics{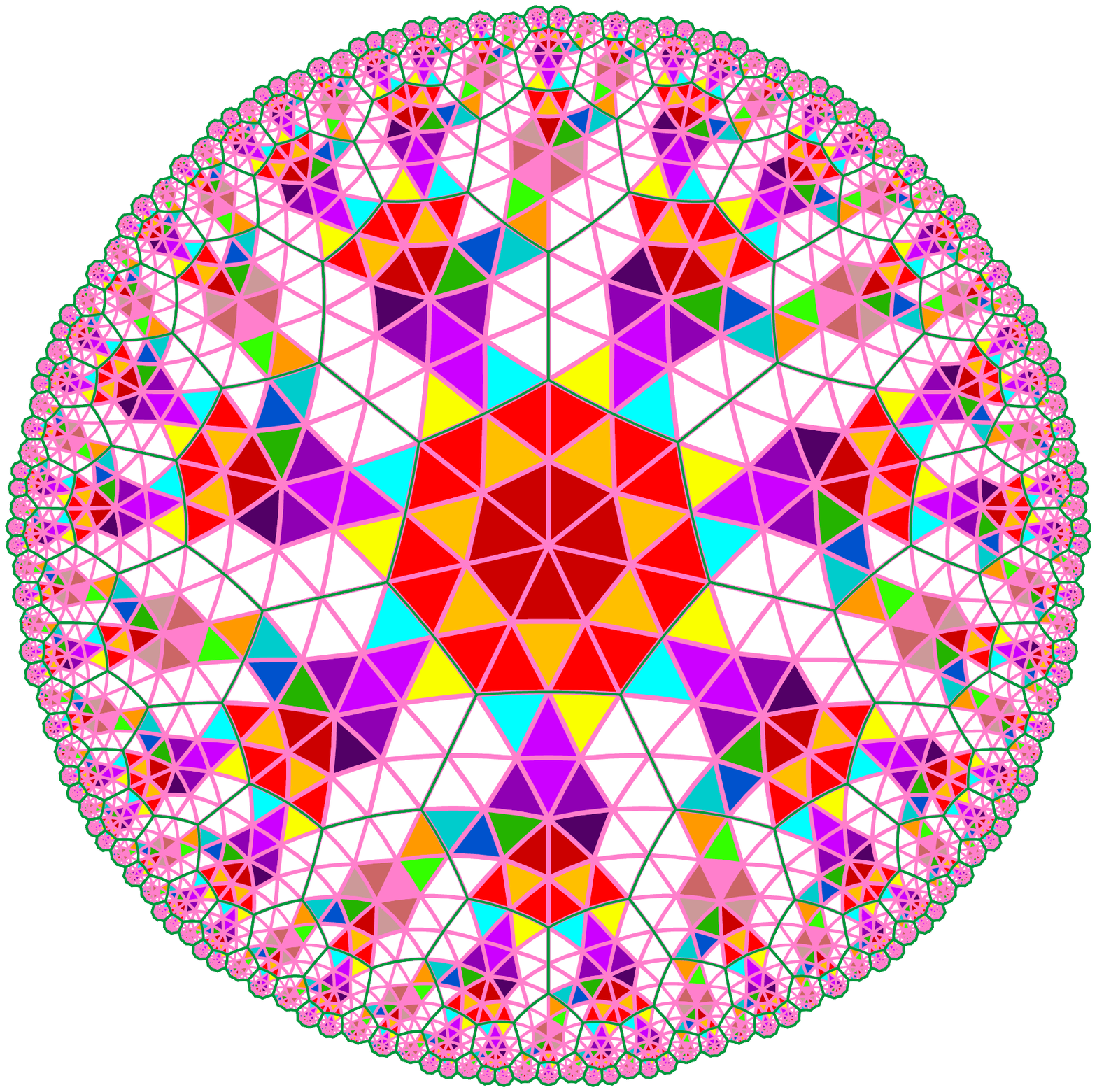}}
\scalebox{0.30}{\includegraphics{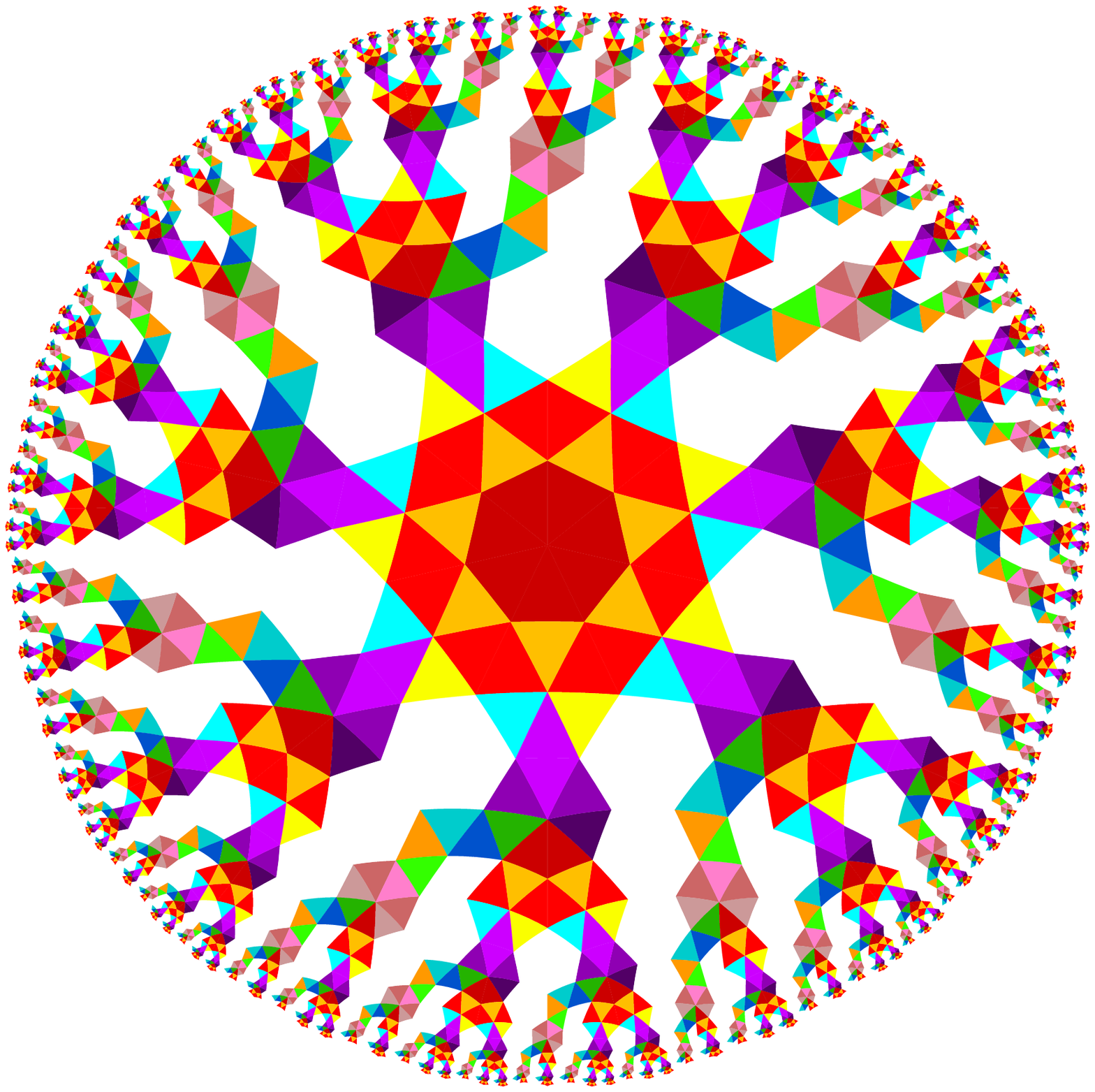}}
\hfill}
\begin{fig}\label{propa_tree_fig}
\leurre
The propagation of the tree structure of the heptagrid. Left-hand side: with the
indication of the grid. Right-hand side: the grid has been removed.
\end{fig}
}

   It seems to us that the result has a striking similarity with pictures
about the growth of colonies of bacteria in highly stressed conditions.

\vspace{-230pt}
\vtop{
\ligne{\hfill
\hskip 20pt\scalebox{0.30}{\includegraphics{propa_bio0.ps}}
\hskip 70pt
\raise 115pt\hbox{\scalebox{0.85}{\includegraphics{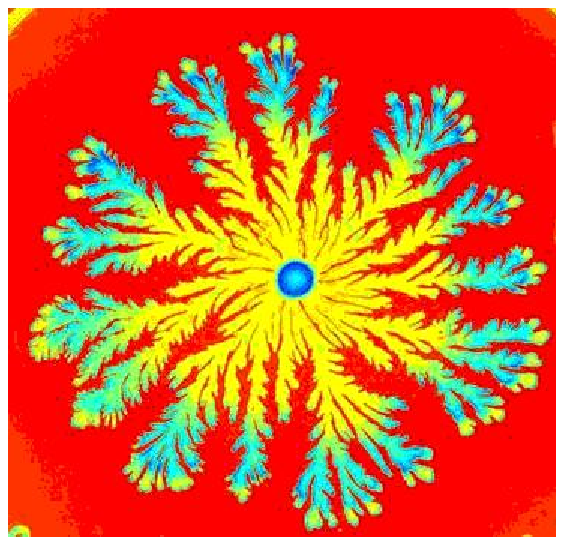}}}
\hfill}
\vspace{-0pt}
\begin{fig}\label{propa_bacteria}
\leurre
Left-hand side: propagation of the tree structure of the heptagrid in the
heptatrigrid. Right-hand side, propagation of a bacteria colony,
picture by courtesy of Professor Ben-Jacob.
\end{fig}
}

We focus on this topic in the next sub-sections trying to reduce the
number of states and to define more acceptable rules from a biologic point
of view.

\subsection{The other simulations}
\label{othersimul}
   In this sub-section, we successively examine three attempts to simulate
the propagation of colonies of bacterias. We shall consider the number of states
we use as well as the information that the cells are assumed to know about 
themselves. We shall try to give the states and these assumptions a kind of
biological flavour. We shall have to keep in mind the specificity of the cellular
automaton programming. In particular, a cell cannot directly act upon another one.
Such an action has to be 2-stepped: the cell~$c$ which wants to act on a 
neighbour~$n$ has to signal the intention of this action by taking a particular
state. Seeing this state on~$c$, and possibly seeing an additional information
displayed by its other neighbours, $n$~can interpret the intention and take the
state {\it wished}, so to say, by its neighbour~$c$. This is why we shall
say that the cell does this and that in our sequel, although, strictly speaking,
the cell can perform such actions in an indirect way, as was just sketched
before.

\subsubsection{With two states}

   We start with a very rudimentary situation. We have two states: white for 
the medium, black for the colony. The cells want to propagate, but competition
is not encouraged. In these conditions the rules can be stated in a simple
way as follows:
\vskip 5pt
\ligne{\hfill
\vtop{\leftskip 0pt\hsize=300pt
$(a)$ \vtop{\leftskip 0pt\parindent 0pt\hsize=260pt\it
A black cell remains black.} 
\vskip 2pt
$(b)$ \vtop{\leftskip 0pt\parindent 0pt\hsize=260pt\it
A white cell becomes black if and only if it has
exactly one black neighbour at this time.
}
}
\hfill
}
\vskip 5pt

From condition $(a)$, once a white cell~$c$ becomes black at time~$t_0$,
it is black for all times~$t$ with $t\geq t_0$.
\vskip 10pt
\vtop{
\ligne{\hfill
\scalebox{0.35}{\includegraphics{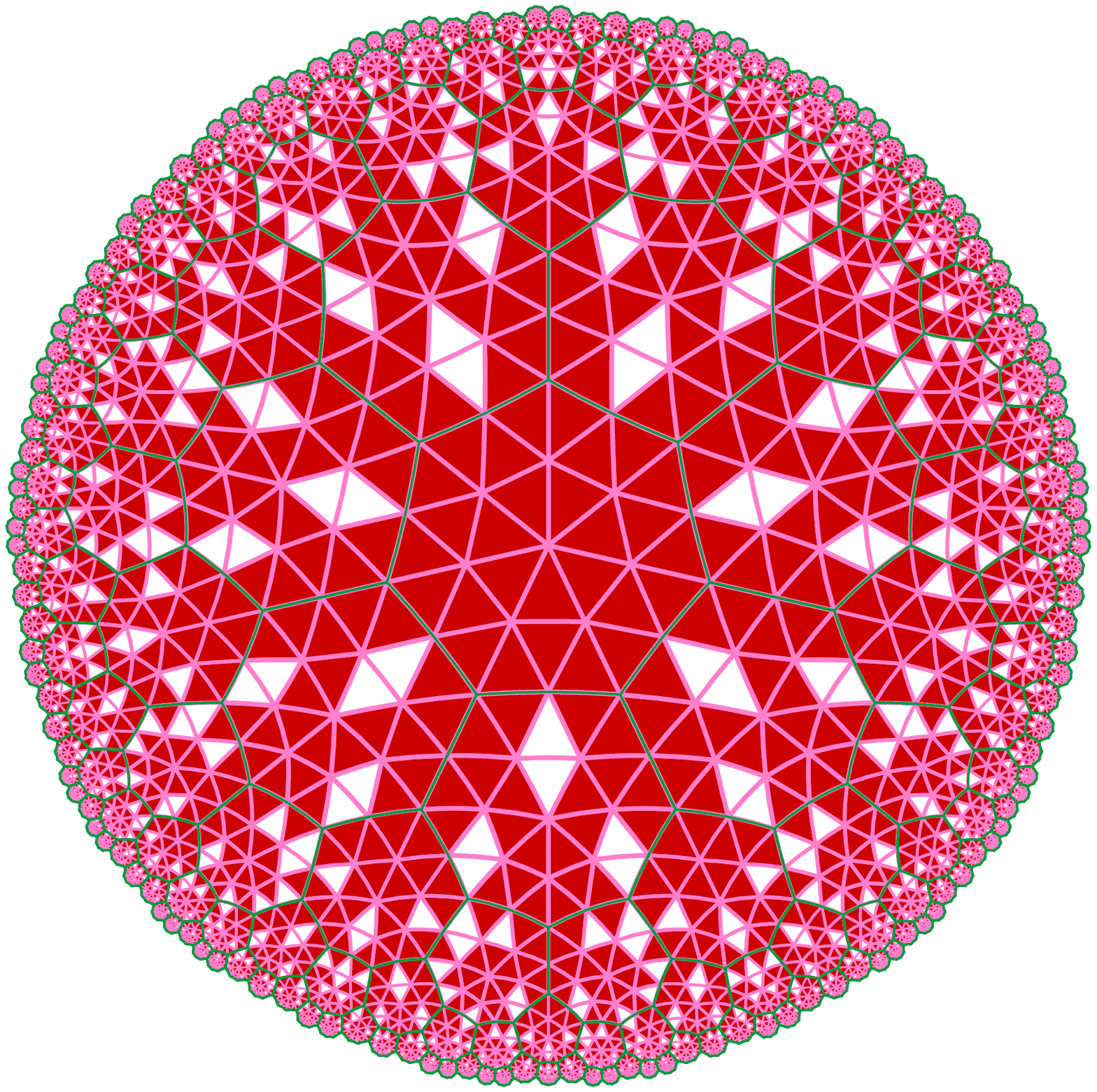}}
\hfill}
\begin{fig}\label{diff2st}
\leurre
Simulation of a diffusion process with $2$~states.
\end{fig}
}

   Figure~\ref{diff2st} illustrates the 36$^{\rm th}$ time of this situation
starting from the core-2 configuration. We can see that the colony invades
almost all the space, leaving holes unoccupied. The condition on the change
of the white cell to a black one has, as a consequence, that a white cell
which has two black neighbours exactly remains white. This the reason
of the pairs of adjacent white cells which are regularly produced in the
evolution of the automaton.

\subsubsection{With four states: version 1}

   In this sub-subsection, we give us more states: four of them, $W$,
$R$, $Y$ and~$V$ calling them white, red, yellow and vermilion respectively. 
As previously, white represents the medium. Red is almost the initial configuration
which is, here again, the core-2 configuration. Then yellow or vermilion can
represent the colony: either one of them or both.

This time too, we formulate the rules in the same style as previously, by properties
on the number of states which are around the cell.

\vskip 7pt
\ligne{\hfill
\vtop{\leftskip 0pt\hsize=300pt
$(a)$ \vtop{\leftskip 0pt\parindent 0pt\hsize=260pt\it
A red, yellow or vermilion cell remains in its colour.} 
\vskip 2pt
$(b)$ \vtop{\leftskip 0pt\parindent 0pt\hsize=260pt\it
A white cell becomes red, yellow or vermilion if and only if, at this time, 
it has exactly one neighbour which is red, yellow or vermilion respectively.
}
}
\hfill}
\vskip 7pt
Here too, when a white cell becomes non-blank, it keeps the new colour for ever.

\vskip 10pt
\vtop{
\ligne{\hfill
\scalebox{0.35}{\includegraphics{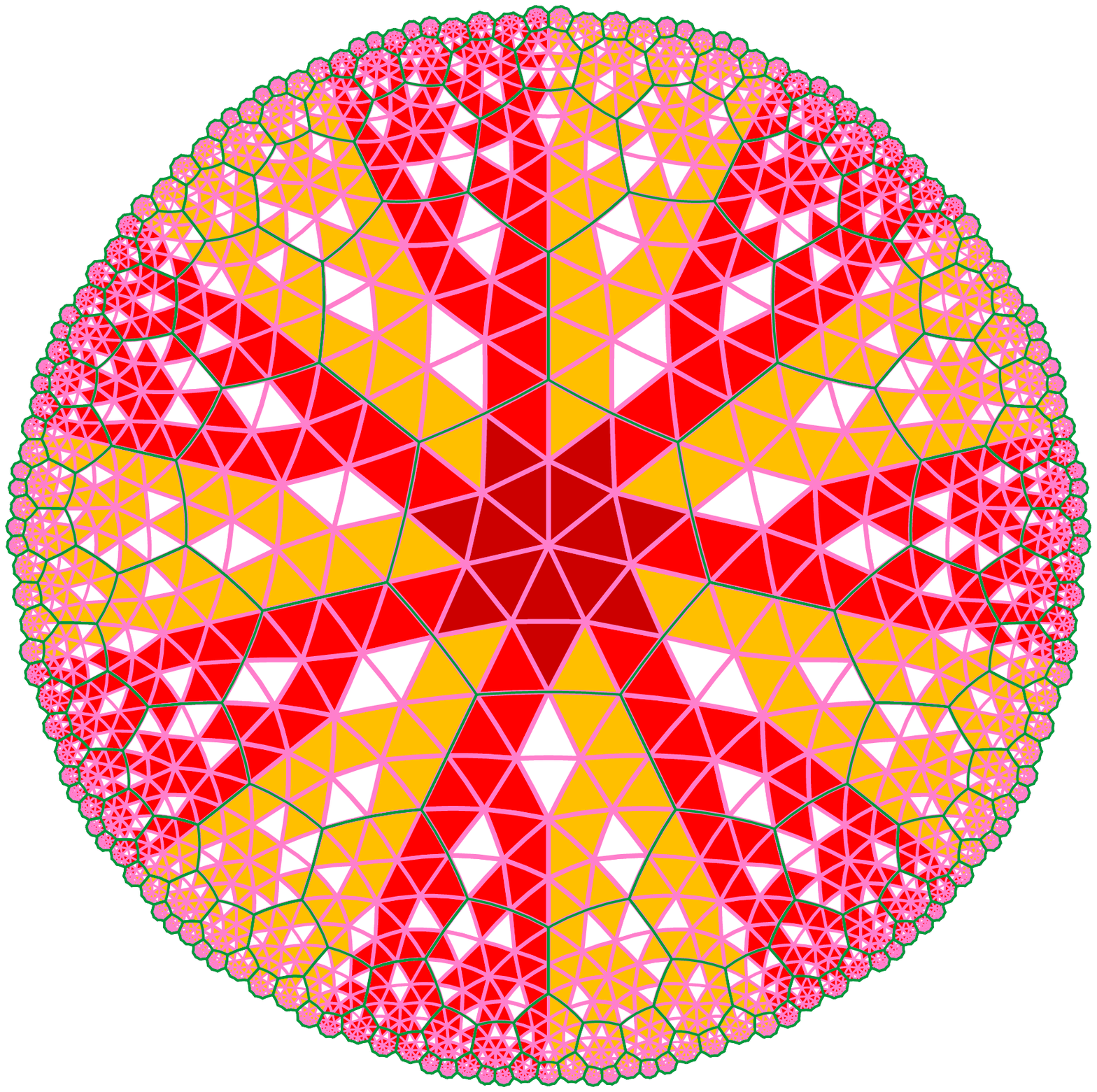}}
\hfill}
\begin{fig}\label{diff4st0}
\leurre
Simulation of a diffusion process with $4$~states, based on a 
local knowledge of the colony.
\end{fig}
}

Figure~\ref{diff4st0} illustrates the 36$^{\rm th}$ step of computation
starting from a configuration where the non white cells occupy a heptagon
exactly, with the pattern we have, in Figure~\ref{diff4st0} for the central
heptagon: the 2-triangles with places 2 or~3 are red, the 2-triangles with
place~0 are vermilion, those with place~1 are yellow. We call this the
heptagonal core configuration. We can notice that in this case also, the
cells which remain white are the same as those of the previous automaton.
We also notice that, thanks to the heptagonal core of the initial configuration,
the red state no more occurs in the computation.  After the initial time,
the computation outside the heptagonal core involves three states only:
the white, the yellow and the vermilion one. 

\vskip 10pt
\vtop{
\ligne{\hfill
\scalebox{0.325}{\includegraphics{diffus4st0.ps}}
\raise-65pt\hbox{\scalebox{0.275}{\includegraphics{branching3.ps}}}
\hfill}
\vspace{-65pt}
\begin{fig}\label{comp_diff4st0}
\leurre
Comparing the diffusion process with $4$~states based on 
local knowledge of the colony with a colony of Figure~{\rm\ref{bacteria}},
picture by courtesy of Professor Ben-Jacob.
\end{fig}
}

   If we look carefully at the computation outside the heptagonal core,
we can see that yellow and vermilion areas are very similar. Each one covers a
larger and larger area. Consequently, we have to reasonably consider that both
of them represent the colony, see Figure~\ref{comp_diff4st0} for
a comparison with one of the colonies illustrated in Figure~\ref{bacteria}. 
We can view them as two sub-species of the colony which do not mix. 

\subsubsection{With four states: version 2}

   In this sub-subsection, we again have four states. However, we have an additional
assumption. We assume that the colony has some knowledge of the geometry of
the space. This can be viewed as an acquired experience of the space by the colony.
The knowledge we assume is that a cell knows its place and whether it is in 
slice~1 or not. We can see that this is a very limited information: 1 bit for the
slice and two ones for the space, which is rather few. We take the same colours
as previously, with white as the state for the space. Here too, the initial 
configuration is the heptagonal core.

This time too, the formulation of the rules is in the same style as previously
but it becomes more intricate, as it involves the place and the slice of
a 2-triangle.

\vskip 7pt
\ligne{\hfill
\vtop{\leftskip 0pt\hsize=300pt
$(a)$ \vtop{\leftskip 0pt\parindent 0pt\hsize=260pt\it
A red, yellow or vermilion cell remains in its colour.} 
\vskip 2pt
}
\hfill}
\ligne{\hfill
\vtop{\leftskip 0pt\hsize=300pt
$(b)$ \vtop{\leftskip 0pt\parindent 0pt\hsize=260pt\it
If a white cell has two white neighbours and if its slice is~$1$,
then it takes the colour of its third neighbour.}
\vskip 2pt
$(c)$ \vtop{\leftskip 0pt\parindent 0pt\hsize=260pt\it
If a white cell has two white neighbours and if its slice is not~$1$
but its place is~$2$ then if its third neighbour is red, yellow or vermilion,
it becomes yellow, vermilion or red respectively.}
\vskip 2pt
$(d)$ \vtop{\leftskip 0pt\parindent 0pt\hsize=260pt\it
If a white cell has two white neighbours and if its third neighbour is red
then, if its place is~$3$, $0$ or~$1$, it becomes red, vermilion or yellow
respectively.}
\vskip 2pt
$(e)$ \vtop{\leftskip 0pt\parindent 0pt\hsize=260pt\it
If a white cell has the states white, yellow and vermilion among its three
neighbours, if its slice is~$1$ and if its place is~$3$, then it becomes red.}
}
\hfill}
\vskip 7pt
With these rules, the cellular automaton behaves in a somewhat different manner.
As can be seen from Figure~\ref{diff4st+}, although the initial configuration 
is the same as previously, the four states are now involved during the whole
computation. Moreover, the colony does not occupy the whole space: the branches
which regularly are spread out are far away from each other, which avoid any kind
of competition. Also, we can see that this time we have a cooperation between the
states. The knowledge whether a cell is in a slice~1 or not allows 
\ligne{\hfill}
\vtop{
\ligne{\hfill
\scalebox{0.35}{\includegraphics{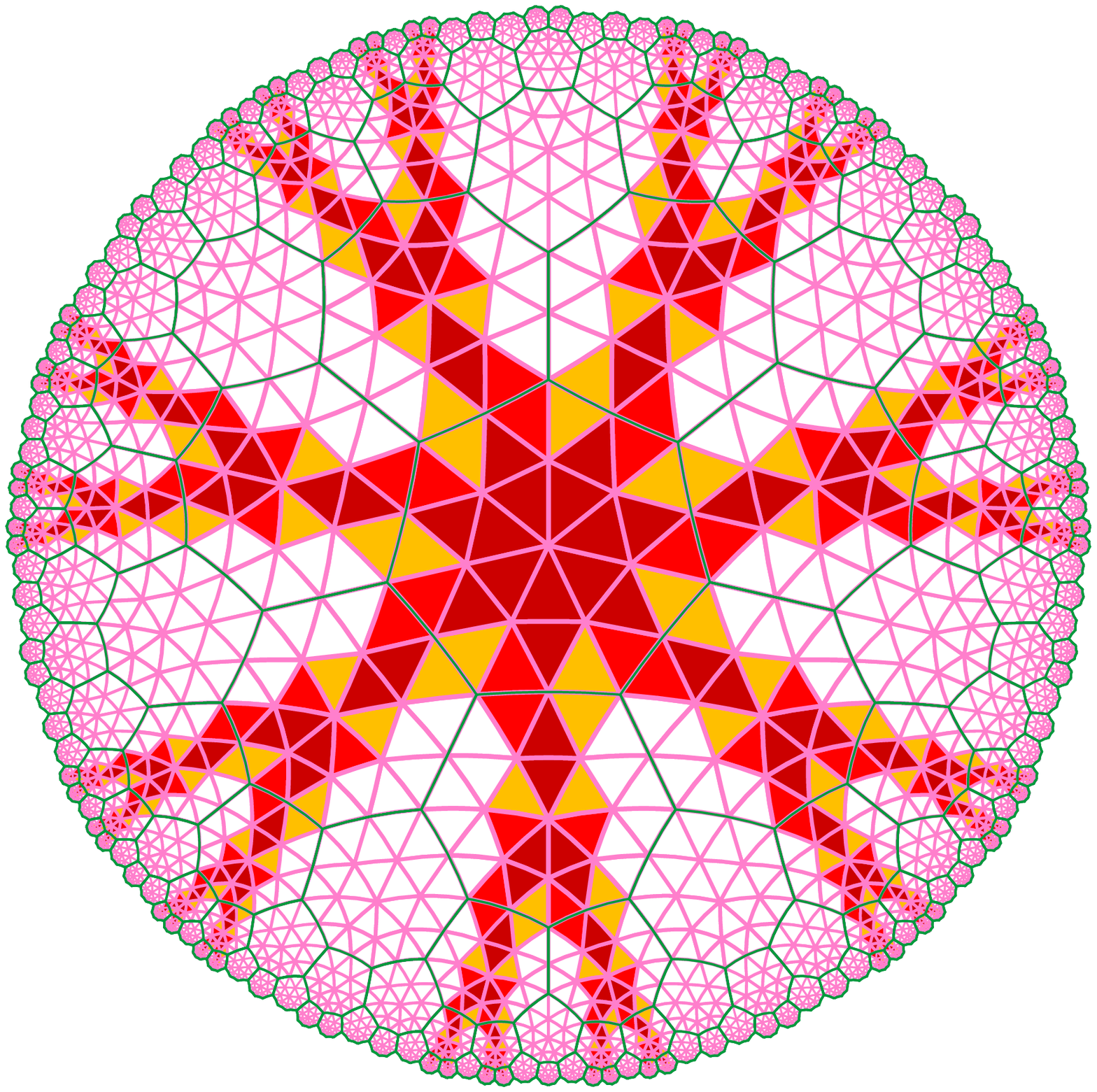}}
\hfill}
\begin{fig}\label{diff4st+}
\leurre
Simulation of a diffusion process with $4$~states, based on a local
knowledge of the colony plus a slight insight on the global structure.
\end{fig}
}

\noindent
the colony
to take advantage of the topology in order to invade the center of a heptagon,
according to the scenario contained in the condition~$c$ of the rule. Next,
the condition~$d$ and~$e$ allow to occupy the slices~4 and~5 of the heptagon
and those ones only, without knowing the number of the slice. This obtained
by the combination of the conditions~$c$, $d$ and~$e$. Once this is checked for
one heptagon around the central cell, this is repeated for all the heptagons
which are the 4- and 5-neighbours of a heptagon. This way, we obtain seven 
binary trees which grow from the heptagonal core. 

   In Figure~\ref{compare_bacteria}, we compare the growth of
Figure~\ref{diff4st+} with a picture of another bacteria colony.
Note that the left-hand side picture of Figure~\ref{compare_bacteria}
is obtained from Figure~\ref{diff4st+} by removing the drawing of the
heptatrigrid. The computer program which draws the figure writes down
a PostScript file from the information obtained by performing the
simulation up to the 36$^{\rm th}$ step, starting from the heptagonal core.
In this writing, the program simply removes
the drawing commands used to draw Figure~\ref{diff4st+}, 
simply keeping the filling commands which allow to paint close areas
defined for drawing the same figure.
 
\vskip 10pt
\vtop{
\vspace{-260pt}
\ligne{\hfill
\hskip 20pt
\scalebox{0.30}{\includegraphics{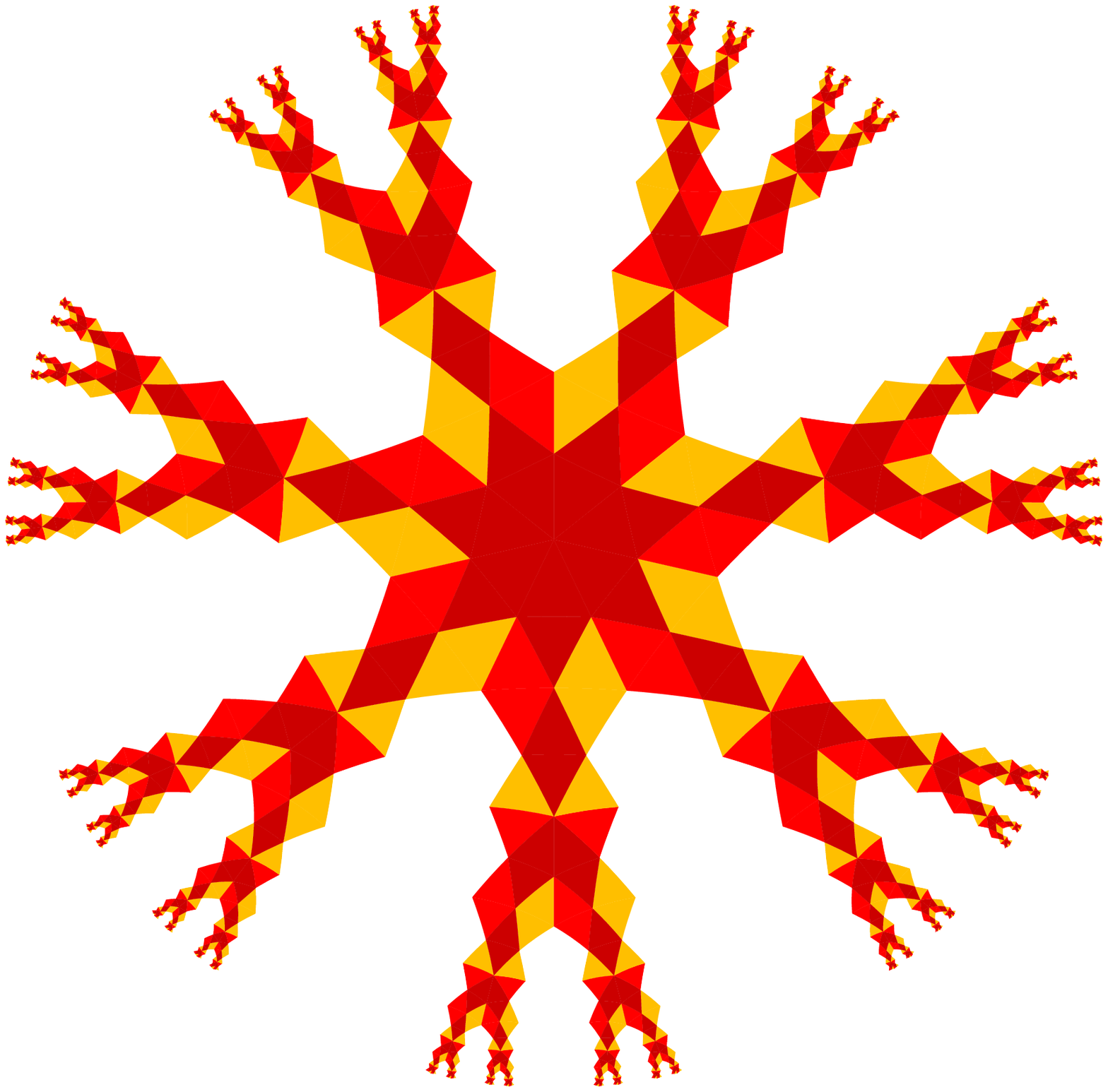}}
\hskip 70pt
\raise 115pt\hbox{\scalebox{0.85}{\includegraphics{bacteria_1.ps}}}
\hfill}
\vspace{-0pt}
\begin{fig}\label{compare_bacteria}
\leurre
Left-hand side: simulation of the growth of a bacteria colony with the
heptatrigrid. Right-hand side, propagation of another bacteria colony,
picture by courtesy of Professor Ben-Jacob.
\end{fig}
}

\section{Conclusion}
\label{conclusion}
   It should be remarked that in all the previous simulations, the computation
may be as long as wished within the time and memory limits of a computer.
Due to the exponential growth of the number of 2-triangles as we go away from
the central heptagon, these limits are rapidly reached and improvements in 
technology may perhaps allow us by one round of 2-triangles further each 
time the capacity is multiplied by~3. However, for simulations of actual 
colonies of bacteria,
this is not a problem as their growth is not only finite but also small 
in the hyperbolic scale.

   It seems to us that this hyperbolic simulation gives an interesting
approximation of the phenomenon observed in real experiments. The above 
discussion
about the space of computation indicates that it could be interesting to
investigate generation~3 of triangles and so, to look at what we obtain
for 3-triangles. Most probably, we could get a finer simulation but certainly
at the price of a bigger number of states. The interpretation of these states
from a biological point of view is of course a question as well as how much of
the knowledge of the space could be allowed for 3-triangles where a third
parameter within the place is necessary.

   These are directions for further work on this topic.

\subsection*{Acknowledgement}
The author is very much in debt to Professor Ben-Jacob for the pictures
he sent him and for the permission of publishing them in this paper.


\begin{thebibliography}{5}
\bibitem{benjacob}
E. Ben-Jacob,
Social behavior of bacteria: from physics to complex organization,
{\it European Physical Journal B}, {\bf 65}(3), (2008), 315-322.

\bibitem{mmkmTCS}
M. Margenstern, K. Morita,
NP problems are tractable in the space of cellular automata in the
hyperbolic plane,
{\it Theoretical Computer Science},
{\bf 259}, (2001), 99-128.

\bibitem{mmJUCSii}
M. Margenstern,
New Tools for Cellular Automata of the Hyperbolic Plane,
{\it Journal of Universal Computer Science},
{\bf 6}(12), (2000), 1226--1252.

\bibitem{mmbook1}
M. Margenstern,
{\it Cellular Automata in Hyperbolic Spaces, vol. $1$, Theory},
Old City Publishing, Philadelphia, (2007), 422p.

\bibitem{mmbook2}
M. Margenstern,
{\it Cellular Automata in Hyperbolic Spaces, vol. $2$, Implementation and
computations},
Old City Publishing, Philadelphia, (2008), 360p.

\bibitem{mmLNCScmc11}
M. Margenstern,
An algorithmic approach to tilings of hyperbolic spaces:
10 years later,
{\it Lecture Notes in Computer Sciences}, {\bf 6501}, (2010), 32-52.

\bibitem{mmTCS4sthepta}
M. Margenstern,
A universal cellular automaton on the heptagrid of the hyperbolic plane with four states,
{\it Theoretical Computer Science}, {\bf 412}, (2011), 33-56

\bibitem{mmsite}
Personal site of M. Margenstern:\vskip 0pt
http://www.lita.univ-metz.fr/\~{}margens/

\end{thebibliography}
\end{document}